\documentclass[10pt,conference]{IEEEtran}
\IEEEoverridecommandlockouts
\usepackage{amsmath,amssymb,amsfonts}
\usepackage{graphicx}
\usepackage{textcomp}
\usepackage{xcolor}
\def\BibTeX{{\rm B\kern-.05em{\sc i\kern-.025em b}\kern-.08em
    T\kern-.1667em\lower.7ex\hbox{E}\kern-.125emX}}

\usepackage[english]{babel}
\usepackage[utf8]{inputenc}

\usepackage{algorithm}   
\usepackage{algpseudocode}
\usepackage{float}
\usepackage{lipsum}
\usepackage{hyperref}
\usepackage{multirow}
\usepackage{amsthm}
\usepackage{array}
\usepackage{scalerel}
\usepackage{bm}
\usepackage{tikz}
\usepackage{csquotes}
\usepackage{diagbox}
\usepackage{gensymb}

\usepackage[T1]{fontenc}

\usetikzlibrary{svg.path}
\usetikzlibrary{calc, fit}

\usepackage[backend=biber, bibstyle=ieee,citestyle=numeric-comp, dashed=false]{biblatex}
\theoremstyle{remark}

\theoremstyle{definition}
\newtheorem{definition}{Definition}
    
\newcolumntype{M}[1]{>{\centering\arraybackslash}m{#1}}

\newcolumntype{L}[1]{>{\raggedright\arraybackslash}m{#1}}

\newcommand{\algstep}[1]{\item[]\medskip\hrule\kern 2pt\hbox to \linewidth{\hspace{\labelsep}\textbf{#1}\hfill}\hrule}

\newlength\myindent
\makeatletter
\newlength{\trianglerightwidth}
\algnewcommand{\LineCommentContAfter}[1]{\Statex \hskip\ALG@tlm%
    \parbox[t]{\dimexpr\linewidth-\ALG@tlm}{\hangindent=\trianglerightwidth \hangafter=1 \strut$\triangleright$ #1\strut}}
\algnewcommand{\LineCommentContBefore}[1]{\Statex \hskip\ALG@thistlm%
    \parbox[t]{\dimexpr\linewidth-\ALG@thistlm}{\hangindent=\trianglerightwidth \hangafter=1 \strut$\triangleright$ #1\strut}}
\makeatother



\makeatletter
\algnewcommand{\NewLineCode}[1]{\Statex \hskip\ALG@thistlm#1}
\algnewcommand{\LineCodeCont}[1]{\State 
    \parbox[t]{\dimexpr\linewidth-\ALG@thistlm}{\hangindent=\trianglerightwidth \hangafter=1 \strut #1\strut}}
\makeatother

\definecolor{orcidlogocol}{HTML}{A6CE39}
\tikzset{
  orcidlogo/.pic={
    \fill[orcidlogocol] svg{M256,128c0,70.7-57.3,128-128,128C57.3,256,0,198.7,0,128C0,57.3,57.3,0,128,0C198.7,0,256,57.3,256,128z};
    \fill[white] svg{M86.3,186.2H70.9V79.1h15.4v48.4V186.2z}
                 svg{M108.9,79.1h41.6c39.6,0,57,28.3,57,53.6c0,27.5-21.5,53.6-56.8,53.6h-41.8V79.1z M124.3,172.4h24.5c34.9,0,42.9-26.5,42.9-39.7c0-21.5-13.7-39.7-43.7-39.7h-23.7V172.4z}
                 svg{M88.7,56.8c0,5.5-4.5,10.1-10.1,10.1c-5.6,0-10.1-4.6-10.1-10.1c0-5.6,4.5-10.1,10.1-10.1C84.2,46.7,88.7,51.3,88.7,56.8z};
  }
}

\newcommand\orcidicon[1]{\href{https://orcid.org/#1}{\mbox{\scalerel*{
\begin{tikzpicture}[yscale=-1,transform shape]
\pic{orcidlogo};
\end{tikzpicture}
}{|}}}}

\IEEEoverridecommandlockouts
\begin{document}

\title{Enhancing Cyber-Resilience in Self-Healing Cyber-Physical Systems with Implicit Guarantees\\
\thanks{}
}

\author{\IEEEauthorblockN{Randolph Loh}
\IEEEauthorblockA{\textit{Cyber Security Strategic Technology Centre} \\
\textit{Singapore Technologies Engineering Ltd}\\
Singapore \\
\orcidicon{0000-0001-8132-4266} \href{https://orcid.org/0000-0001-8132-4266}{0000-0001-8132-4266}}
\and
\IEEEauthorblockN{Vrizlynn L. L. Thing}
\IEEEauthorblockA{\textit{Cyber Security Strategic Technology Centre} \\
\textit{Singapore Technologies Engineering Ltd}\\
Singapore \\
\orcidicon{0000-0003-4424-8596} \href{https://orcid.org/0000-0003-4424-8596}{0000-0003-4424-8596}}
}

\maketitle

\begin{abstract} 
Self-Healing Cyber-Physical Systems (SH-CPS) effectively recover from system perceived failures without human intervention. They ensure a level of resilience and tolerance to unforeseen situations that arise from intrinsic system and component degradation, errors, or malicious attacks. 
Implicit redundancy can be exploited in SH-CPS to structurally adapt without the need to explicitly duplicate components. However, implicitly redundant components do not guarantee the same level of dependability as the primary component used to provide for a given function. Additional processes are needed to restore critical system functionalities as desired. 
This work introduces implicit guarantees to ensure the dependability of implicitly redundant components and processes. Implicit guarantees can be obtained through inheritance and decomposition. Therefore, a level of dependability can be guaranteed in SH-CPS after adaptation and recovery while complying with requirements. We demonstrate compliance with the requirement guarantees while ensuring resilience in SH-CPS.
\end{abstract}

\begin{IEEEkeywords}
Self-Healing, Structural Adaptation, Redundancy, Guarantees, CPS, Resilience
\end{IEEEkeywords}

%%%%%%%%%%%%%%%%%%%%%%%%%%%%%%%%%%%%%%%%%%%%%%%%%%%%%%%%%%%%%%%%%%%%%%%%
% intro 
\section{Introduction}\label{sect:1}
Self-healing (SH) systems detect and recover from faults and failures without human intervention \cite{GHOSH20072164}. They monitor themselves and their environments to assist decision-making processes when selecting and deploying recovery plans when faults and failures occur. SH systems self-diagnose to determine when a fault has occurred, the type of fault, and its severity \cite{Ma2018ModelingFF}. SH systems decide on the actions required to adapt to the fault and recover, referring to a knowledge base that incorporates user experiences and historical data. Adaptive systems observe the MAPE-K feedback control loop, where structures of autonomic elements operate in sequence to \emph{Monitor} and \emph{Analyse} events, \emph{Plan} and \emph{Execute} actions necessary to adapt while referencing a shared \emph{Knowledge base} \cite{1160055,7194653}. 
Systems can implement redundancy in static, dynamic, or hybrid configurations to achieve fault tolerance \cite{Nelson1990FaulttolerantCF}. Such systems can also be categorised as explicitly or implicitly redundant, the former being more costly \cite{8606923}. Explicitly redundant systems feature the use of multiple identical components, whereas implicitly redundant systems factor distinct components with similar but varying degrees of capabilities and support different primary functions. 
Explicitly redundant systems have operational considerations that are replicated across identical components. This also propagates risks across identical backup components and more so for malicious faults. This is absent in implicitly redundant systems. The inherently different components found in implicitly redundant systems reproduce functionality of other components with additional processes. This makes it difficult for faults to propagate across components though such systems may perform differently after recovery. 

Components in Cyber-Physical Systems (CPS) collectively make smart decisions to function as intended \cite{5523280,8606923}. CPS should maintain a level of dependability even as faults and failures occur \cite{Miclea2011AboutDI}. 
Events affecting the dependability of a CPS can negatively impact the system and its users. Therefore, SH capabilities are desirable for CPS to detect, mitigate, and recover from faults and failures. SH-CPS has garnered considerable interest in recent years, particularly in autonomous systems, as need for human management and mediation is minimised \cite{s19051033}. Nonetheless, ensuring the dependability of such systems is a challenge due to increasing complexity. 

This work examines dependability guarantees of implicitly redundant SH-CPS. Explicitly redundant SH-CPSs can assume identical components adopt identical guarantees, thereby maintaining an equal level of dependability after recovery. 
This is not applicable in implicitly redundant SH-CPS as the underlying components are inherently different and do not guarantee the same level of dependability. Furthermore, additional processes are often required when adopting implicitly redundant components to provide similar functionalities and services. 

% state contributions
The main contributions of this work are as follows.
\begin{itemize}
\item We introduced \emph{implicit guarantees} for SH-CPS that leverages implicit redundancy. Guarantees can be \emph{inherited} or \emph{decomposed} to establish a minimum required level of dependability for SH-CPS.
\item We examine considerations when \emph{inheriting} guarantees from the preceding components and \emph{decomposing} guarantees from succeeding components for implicit guarantees. 
\item We implemented, demonstrated, and evaluated our work on existing self-adaptation approaches, such that SH-CPS factor implicit guarantees during recovery. 
\end{itemize}

% paper structure
This paper is organised as follows. Related works are presented in Section~\ref{sect:2}. Section~\ref{sect:3} specifies information to guarantee the reliability in SH-CPS. Section~\ref{sect:4} details implementation, experiments, and analysis. Additional concerns are shared in Sections~\ref{sect:5} and~\ref{sect:6} before concluding in Section~\ref{sect:7}.

%%%%%%%%%%%%%%%%%%%%%%%%%%%%%%%%%%%%%%%%%%%%%%%%%%%%%%%%%%%%%%%%%%%%%%%%
% related work 
\section{Related works}\label{sect:2}

\subsection{Self-healing and self-adaptation}
SH systems can discover, diagnose, react, and recover from system perceived faults and failures to minimise disruptions and ensure continuity \cite{5386835}. 
Adaptive systems constantly monitor themselves and the environment to determine if there is a need to modify their behaviour to deal with unpredictability \cite{Musil2017}. 
A mixed-integer programming approach that tries to maximise a fitness function to find the optimal service and component composition to provide high quality of service (QoS) was presented in \cite{Wang2020ServiceCI}. 
A multi-agent task-orientated functional communications approach aims to preserve properties of distributed systems was proposed in \cite{Rajput2021MultiagentAF}. 
Authors demonstrated system recovery by triggering recovery actions observing timings between events when a fault occurs in \cite{Abdi2018GuaranteedPS}. 
Adaptation can be categorised as parametric and structural \cite{Cheng2009}. The former modifies system parameters to manipulate behaviour, the latter is concerned with the inter-connectivity of components. 
Uncertainties brought about by the environment, availability of resources, and goals of users affect the reliability, behaviour, and also impact critical operations of SH-CPS \cite{weyns2017software}. They are often modelled to evaluate self-healing, fault correction, and fault tolerance capabilities of the CPS in unforeseen situations \cite{10.1007/978-3-319-42061-5_16,Zhang2019,8606923,Musil2017,9466159}. 

\subsection{Redundancy in self-healing systems}
Redundancy is a practical solution for systems to improve resilience and ensure a level of dependability. 
Explicitly redundant systems are often expensive due to the use of duplicate components. 
Modular redundancy explicitly uses interchangeable components to detect faults and support voting mechanisms \cite{8606923,KHALIL2019104620}. 
Where a framework demonstrated a 2-out-of-3 circuit voting mechanism leveraging redundant components to verify operations and recover to a good state in \cite{10.1007/978-3-319-58469-0_4}. 
Implicitly redundant systems offer similar features without duplicate components. 
In \cite{1432851}, the authors leverage causal relations of sensors to establish validity levels to validate sensor values and implicit sensor redundancy through a sensor redundancy graph. 
A temporally predicable framework for re-configurable embedded real-time systems was presented in \cite{6913205}. A service orientated approach that searches the system ontology for semantically equivalent services as substitute for the affected service. 
Structural adapting approaches leveraging implicitly redundant components within a CPS to recover from failure were presented in \cite{8387659}. 
The authors compared methods including Ontology-based Run-time Reconfiguration (ORR) \cite{6913205}, Depth-First Search (DFS), and Self-Healing by Property-Guided Structural Adaptation (SHPGSA) \cite{8387659}. 
None of the aforementioned approaches considered the level of dependability of the returned solutions. 

\subsection{Guaranteeing dependability}
CPS should guarantee a level of dependability. Especially when systems are affected as the infrastructure adapts during operation \cite{10.1145/3424771.3424804}. 
Components within the CPS also guarantee performance as specified by the manufacturers, users, and underlying applications. 
Functional correctness should be guaranteed alongside adaptation goals as more works gather increased interest in the security, guarantee and verification of self-adaptive systems \cite{WONG2022106934,weyns2017software}. 
Simplex Control Adaptation (SimCA) tries to comply with multiple requirement goals to provide guarantees by first identifying dependencies within the system, then synthesises a set of controllers, before finally carrying out operations that control and optimise the system goals \cite{10.1145/2950290.2950301}. 
A five-step process that ingests several user inputs to automatically synthesise a controller to manage trade-offs between multiple goals in order to provide stronger guarantees was described in \cite{10.1145/3106237.3106247}. 
The ENgineering of TRUstworthy Self-adaptive sofTware (ENTRUST) methodology was introduced to develop trustworthy self-adaptive software and assurance cases \cite{8008800}. ENTRUST incorporates modelling and verification into a seven staged industry adopted assurance process. 
While identical components in explicitly redundant systems provide the same guarantees, implicitly redundant systems with comparable components provide different guarantees. 
Moreover, intermediate components that process data may be exposed to uncertainties. 
These proxies may carryover similar guarantees but may not ensure dependability. This is common in systems with composite services to provide a level of service and reliability \cite{Wang2020ServiceCI}. 
It is necessary to establish guarantees for these proxies before determining how a SH-CPS recovers. SH-CPS should provide a minimum level of dependability after recovering from a fault. We refer to this as \emph{implicit guarantees}.

%%%%%%%%%%%%%%%%%%%%%%%%%%%%%%%%%%%%%%%%%%%%%%%%%%%%%%%%%%%%%%%%%%%%%%%%
% implicitly redundant properties and guarantees
\section{Self-healing, redundancy and guarantees}\label{sect:3}
A SH-CPS is a heterogeneous system comprising hardware and software. 
The SH-CPS structurally adapts when a failure or fault is detected owing to a knowledge base that contains information on implicitly redundant components. The knowledge base is searched for substitutes to restores the system to a functional state. 
\emph{Implicit guarantees} are introduced to ensure dependability in implicitly redundant systems and to support the SH-CPS adaptation process to source solutions that guarantee a level of dependability. 

\subsection{System model}\label{subsec:1a}
A CPS is composed of functional blocks defined by processes and the flow of information \cite{s21062140}. 
The CPS $(Z)$ consists of subsets of interconnected cyber-physical components $(z)$ that may be dynamically reconfigured. Functional information required by the system to perform its tasks is transmitted between components, where each component is associated with \emph{functions} $(f)$ that generate \emph{outputs} $(O)$ given \emph{inputs} $(I)$. 
A \emph{failure} occurs when a component deviates from its normal operation and may affect multiple interrelated components. A monitor observes the state of the system and decides when to trigger a recovery action \cite{GHOSH20072164}. 
Irregularities and deviations are identified by comparing behavioural differences in space and time \cite{8780540}. 
A smart car may have multiple sensors that read similar parameters such that a faulty sensor may be substituted to guarantee some level of functionality. 
This indicates the presence of \emph{implicit redundancy}. 

\begin{figure}[htbp]
\centering
\includegraphics[height=4cm]{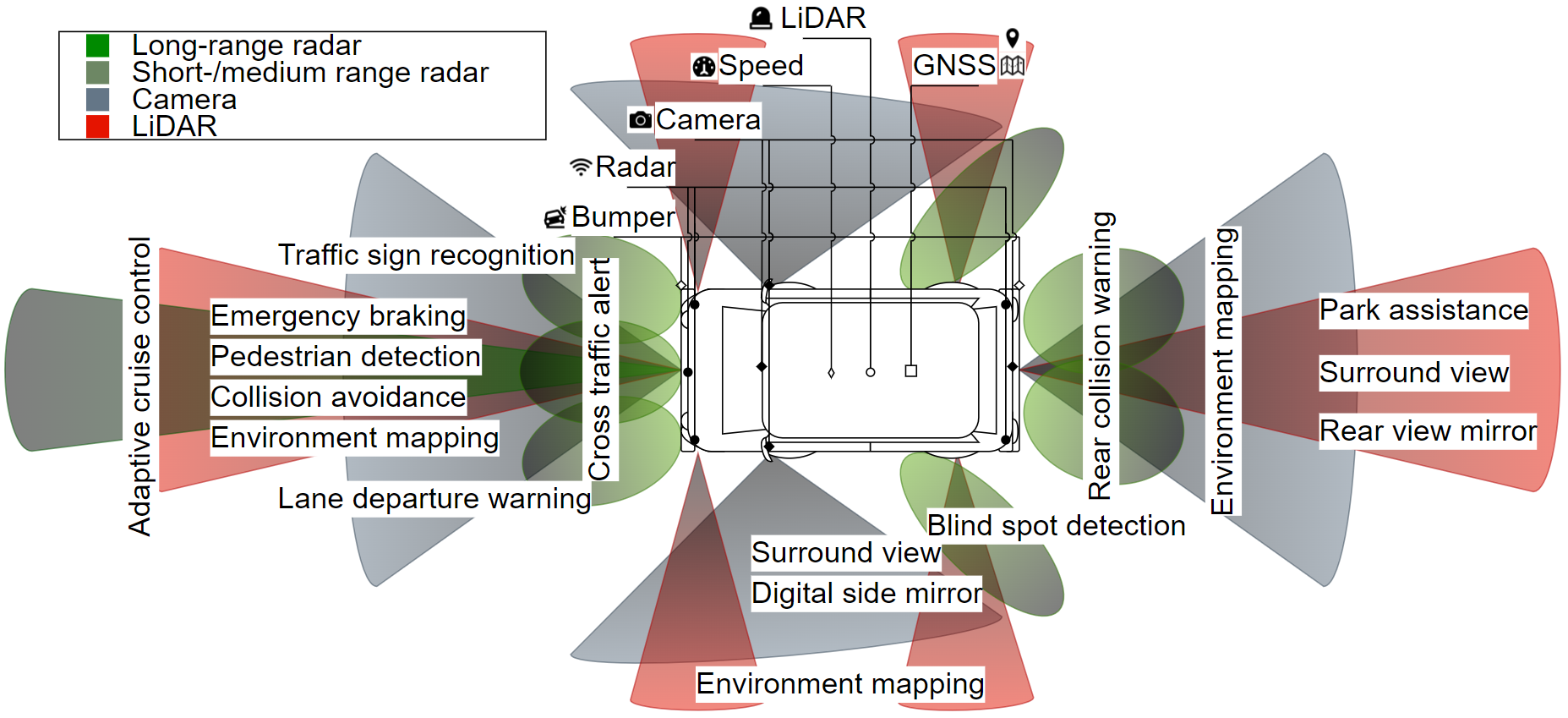}
\caption{An autonomous vehicle and sensors that allow it to make sense of its environment. The different sensors monitor different areas to enable various applications, as depicted in \cite{s21062140}}
\label{fig:architecture3}
\end{figure}

% define implicit redundancy
\begin{definition}[Implicit redundancy]\label{def:impRed}
Implicit redundancy is present in system $Z$ when it contains two distinct components $z_x$ and $z_y$ and functions $f_x$ and $f_y$ that can generate the same output $O$ given different inputs $I_x$ and $I_y$.
\end{definition}

\begin{equation}
% \scriptsize
    \begin{split}
        z_x = (f_x,I_x,O) \\
        z_y = (f_y,I_y,O) \\
        z_x,z_y \in Z
    \end{split}
\end{equation}

\subsection{Knowledge base}
\begin{figure}[htbp]
\centering
\includegraphics[height=8cm]{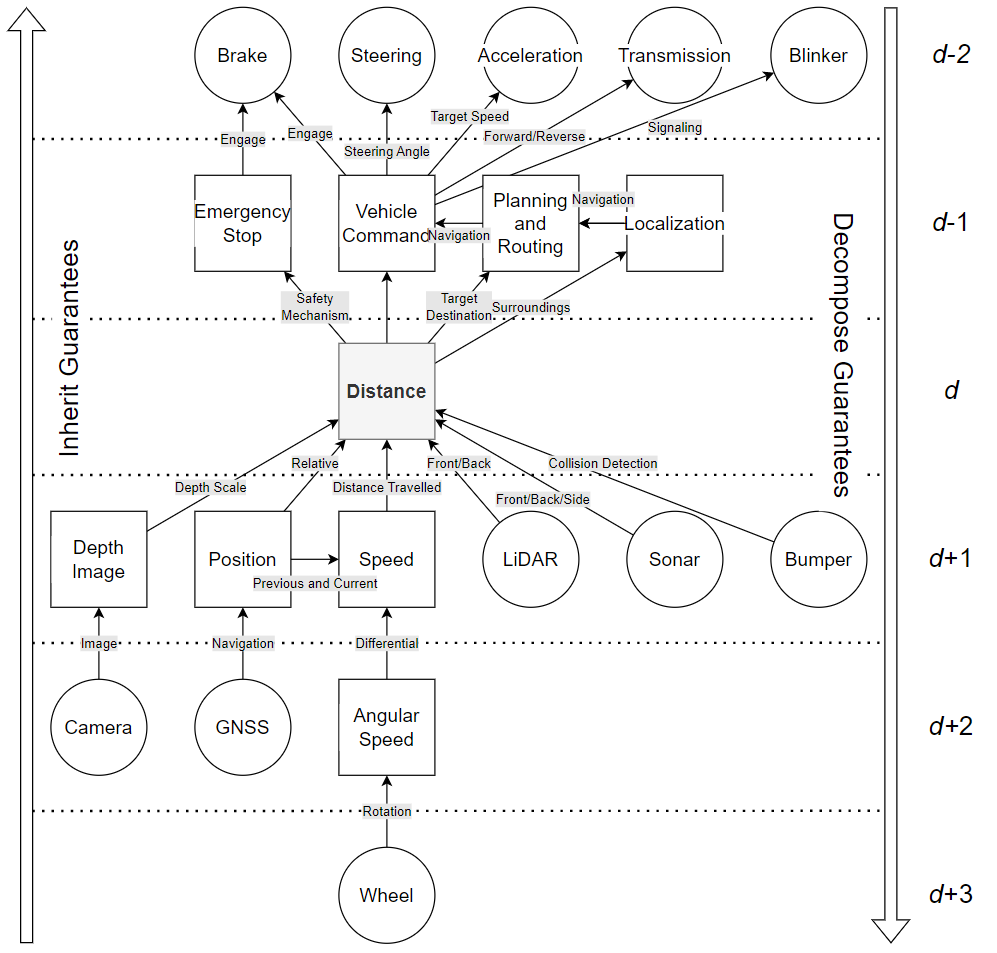}
\caption{Relationships of components and parameters, circle and square shaped nodes. \emph{Distance} related parameters are obtained from various nodes to support a variety of functions. Guarantees are inherited or decomposed.}
\label{fig:knowledgeBase2}
\end{figure}

Information concerning the SH-CPS components, programs and their relationships constitute a knowledge base. The knowledge base $\text{K}=(N,E)$ is represented as a directed graph of hierarchically interconnected nodes where edges $E$ define the relationship between a node $n_{d} \in N$ and its relative set of predecessor nodes $Pred_K(n_{d})$ or successor nodes $Succ_K(n_{d})$. 
Nodes represent components and programs within the SH-CPS, and the edges define relationships as inputs and outputs. These relationships establish a hierarchy between the nodes, as shown in Figure~\ref{fig:knowledgeBase2}. 
Combinations of nodes $N_{c}$ can be generated so that variations of node sequences can be formed from node $n_d$. The nodes within the sequence is established on relationships between predecessor and successor nodes. 

\begin{equation}
% \scriptsize
    Pred_K(n_{d})=N_{d+1}
\end{equation}

\subsection{Properties}
Properties are defined by manufacturers, users, or applications. Components may consist of different properties that describe its operation and environment boundaries. 
The SH-CPS can identify components that satisfy the properties required to restore system functions during recovery. Properties may be influenced by the environment or the component's condition. 

\begin{equation}
% \scriptsize
    P(n) = (p_1,p_2, \cdots ,p_k)
\end{equation}

\subsection{Substitution}
A set of \emph{substitutions} $S_{n_s}$ can be obtained from the knowledge base as potential replacements for a component $n_s$ while satisfying some property requirements.  
Each substitution $s$ is an acyclic sub-graph derived from a subset of nodes and edges within the knowledge base with $n_s$ as the root node. 
Validity of substitutions depends on the availability of all associated components and are necessary for a SH-CPS to continue its operations after a fault. 
Properties are prioritised so that only those necessary are considered during substitution.

\begin{equation}\label{eq:valSub}
% \scriptsize
    \begin{split}
        S_{n_s} = \{s_1,s_2, \cdots ,s_k\} \\
        s = (n_s,N_s,E_s)
    \end{split}
\end{equation}

\subsection{Utility}\label{subsec:1e}
From decision theory, a multi-attribute utility function selects the substitution that satisfies weighted properties required by the system. 
The function $u$ determines the utility of node $n$ as $u_n$ (\ref{eq:multiAttUtil}). This enables the SH-CPS to measure, assess, and rank the performance of substitutions to select the best substitution through normalised utility with values closer to 1 potentially offering better performance. 
A multiplicative utility aggregation function is adapted from \cite{8387659} to calculate utility as a product of nodes in the substitution (\ref{eq:utilCom}). The utility of sequences of nodes decreases monotonically, thus ranking substitutions hierarchically. The best substitution $s_{best}$ is the $argmax$ from a set of valid substitutions $S_{n_{s}}$.

\begin{equation}\label{eq:multiAttUtil}
% \scriptsize
    u_n = u(P(n)) = \sum_{i=1}^k w_iu(p_i), 1 \geq u_n \geq 0
\end{equation}

\begin{equation}\label{eq:utilCom}
% \scriptsize
    u_{N_c} = \prod_{n \in N_{c}} u_n
\end{equation}

\subsection{Implicit Guarantees}
Subsystems and components within a CPS will guarantee some dependability attributed to their properties. 
As additional processes are assumed necessary for two distinct components to produce the same output in implicitly redundant systems, implicitly redundant components and processes may not possess properties required to guarantee dependability. SH-CPS should still ensure a level of dependability after recovery. This work accomplishes this through \emph{implicit guarantees} where implicitly redundant components inherit or pass on some level of guarantees. 
While guarantees should consider individual properties to ensure dependability, this work presents guarantees as utility for simplicity.

\begin{definition}[Implicit guarantees]\label{def:impGua}
An implicitly redundant node $n_d$ in the knowledge base $\text{K}$ guarantees properties $P(n_d)$ inherited from the predecessor nodes of depth $N_{d+1}$ or passed on to the successor nodes of depth $N_{d-1}$. 
\end{definition}

\begin{equation}\label{eq:guaPro}
        n_d.g = P(n_d) = (p_1,p_2, \cdots ,p_k)
\end{equation}

Two approaches are defined to obtain implicit guarantees for an implicitly redundant node (IRN). The first, a bottom-up approach where IRN \emph{inherit guarantees} from predecessor nodes. This identifies properties that the IRN can provide. The second, a top-down approach which successor nodes \emph{decompose guarantees} to the IRN. This identifies properties the IRN must satisfy. Both approaches establishes requirements for the SH-CPS to fulfil during recovery. Therefore, the best substitution $s_{best}$ to replace a failed component $n_s$ from a set of substitutions $S_{n_s}$ should also satisfy guarantee requirements of the IRN in the SH process. 

\begin{equation}
        s_{best} = argmax_{s \in S_{n_s}} u_s, \text{if } u_s \geq n_s.g %\label{eq:argMax},
\end{equation}

\begin{definition}[Inherit guarantees]\label{def:inhGua}
Implicitly redundant nodes inherit some level of guarantees from predecessor nodes. Inherited guarantees describe the level of dependability that the implicitly redundant node is able to achieve from guarantees of predecessor nodes. 
\end{definition}

\begin{definition}[Decompose guarantees]\label{def:decGua}
Some level of guarantees from successor nodes is decomposed to implicitly redundant nodes. Decomposed guarantees represent a level of functionality that the implicitly redundant node must provide to guarantee a level of dependability. 
\end{definition}

\begin{algorithm}
    % \scriptsize
    \caption{$InheritGuarantee(n_i)$} \label{alg:1}
    \begin{algorithmic}[1]
        \renewcommand{\algorithmicrequire}{\textbf{Require:}}
        \renewcommand{\algorithmicensure}{\textbf{Ensure:}}
        \Require $|Pred_\text{K}(n_i)|\ge0$, $!n_i.g$
        \State $\textbf{new } propertyList$ 
        \If{$P(n_i)$}
            \State $propertyList.\text{append}(P(n_i))$
        \Else
            \For{$n_{j} \textbf{ in } Pred_\text{K}(n_i)$}
                \State $InheritGuarantee(n_{j})$
                \State $propertyList.\text{append}(n_{j}.g)$
            \EndFor
        \EndIf
        \State $n_i.g \gets argmin(propertyList)$ \label{alg:line:1L10}
    \end{algorithmic}
\end{algorithm}

\subsubsection{Inheriting guarantees} 
In general, IRN inherited properties from their predecessor nodes, hierarchically guaranteeing some level of dependability. Algorithm~\ref{alg:1} describes an iterative process that recursively examines predecessor nodes for properties to inherit. The algorithm performs a DFS of the knowledge base, eventually reaching the leaf node. First, it determines if properties are defined for a given node. The absence of properties indicates the need to traverse the knowledge base. It then attempts to derive the properties of the predecessor nodes, effectively inheriting its properties. This guarantees a level of dependability that references predecessor nodes. The number of predecessor nodes is assumed to be greater than or equal to $0$. Specific to inheriting guarantees, the $argmin$ of each property within a list of properties is preferred (alg:~\ref{alg:line:1L10}). 
The algorithm can also be applied to decompose guarantees from successor nodes $Succ_\text{K}(n_1)$. However, additional factors should be considered when selecting guarantees.

\subsubsection{Selecting guarantees}
System functions require specific properties to operate to guarantee dependability \cite{s21165397}. IRN should guarantee similar levels of dependability by inheriting the same properties. The multi-attribute utility function can be leveraged to select the appropriate properties for the IRN to inherit and achieve a level of dependability. 
Table~\ref{tab:u1} demonstrates the function to rank sensors by relevant properties. 
TurtleBot3 has multiple sensors with distance sensing capabilities \cite{turtlebot3}. They include the 360 Laser Distance Sensor LDS-02, ultrasonic distance sensor HC-SR04, and the Intel RealSense R200 camera. Weights are assigned to properties by their influence on distance sensing functions. They include effective range (Range), accuracy (Acc), field of view (HFOV), polling frequency (Freq), and frames per second (FPS). Values are normalised against 10m, 100\%, 360\degree, 500Hz, and 30FPS respectively. 
Normalising properties is subject to the requirements. For example, a system may require an effective range of 100m. 
Although the minimum level of dependability an IRN can guarantee is identified when inheriting guarantees, an IRN may select properties to satisfy decomposed guarantees such that substitutions must satisfy a level of dependability. 

\subsubsection{Guarantees in dynamic environments}
SH-CPS is also subject to its environment including being susceptible to weather conditions \cite{s21165397}. Performance guarantees are normally for ideal environments, but do not hold in dynamic environments. The unpredictability of the weather can adversely affect the SH-CPS on varying degrees. 
Therefore, SH-CPS should also account for changes in its environment. While predicting changes in the environment is no easy task, SH-CPS can adaptively monitor changes from uncertainties in space and time \cite{8780540}. Effects on guarantees will be explored in future.

\begin{table}[htbp]
\scriptsize
\caption{Guarantee a level of dependability for the distance metric through a multi-attribute utility function.}
\centering
\begin{center}
\begin{tabular}{|M{0.12\linewidth}|M{0.07\linewidth}|M{0.05\linewidth}|M{0.08\linewidth}|M{0.06\linewidth}|M{0.05\linewidth}|M{0.06\linewidth}|M{0.05\linewidth}|}
\hline
$w$ & 0.3 & 0.4 & 0.2 & 0.1 & 0.1 & $W$=1 & \\
\cline{1-7}
\diagbox{$n$}{$p$} & Range (m) & Acc (\%) & HFOV (\degree) &  Freq (Hz) & FPS & Util Score & Rank\\
\hline
LiDAR & 8 & 97 & 360 & 2.3k & - & 0.93 & 1 \\
\hline
Ultrasonic & 4 & 95 & 21 & 40 & - & 0.55 & 3 \\
\hline
Camera & 4 & 97.5 & 70 & - & 60 & 0.65 & 2 \\
\hline
\hline
Min & 4 & 95 & 21 & 40 & 60 & 0.55 & - \\
\hline
Avg & 5.33 & 96.5 & 150.33 & 1.17k & 60 & 0.73 & - \\
\hline
Max & 8 & 97.5 & 360 & 2.3k & 60 & 0.93 & - \\
\hline
\end{tabular}
\label{tab:u1}
\end{center}
\end{table}

%%%%%%%%%%%%%%%%%%%%%%%%%%%%%%%%%%%%%%%%%%%%%%%%%%%%%%%%%%%%%%%%%%%%%%%%
% implementation 
\section{Implementation, experiments, and analysis}\label{sect:4}

Our work was implemented in Python 3.6 on a desktop system with a i7-9700K CPU and 64GB RAM running Ubuntu 20.04. Furthermore, the SH approaches described in \cite{8387659} were modified to also consider guarantee requirements when searching for valid substitutions. 
Approaches include ORR, SHPGSA, and DFS alongside modified counterparts ORRG, SHPGSAG, and DFSG. 
For simplicity, the utility value is observed instead of individual properties. A higher utility value implies a better substitution. We examined the time required to find valid substitutions with the Python in-built time module, the number, and length of returned substitutions. The results were averaged from 100 experiments. 

\subsection{Implementation}
Algorithm~\ref{alg:1} was implemented to traverse the entire knowledge base to assign guarantees starting from the leaf nodes. 
Each SH approach then examines the knowledge base to find valid substitutions. 
They assess substitutions differently as a result of their underlying mechanisms. 
 \emph{ORRG} will repetitively try to look for substitutions until it satisfies guarantee requirements. Measures were needed to prevent the approach from reconstructing the prior substitutions. The returned substitution may not be the optimal solution.
In \emph{SHPGSAG}, workers construct substitutions. Checks were carried out on each worker to verify if it satisfies the guarantee requirements. Workers that did not satisfy the guarantee requirements were immediately aborted even when construction was not completed. 
In \emph{DFSG}, checks were performed during the search to remove substitutions that did not satisfy guarantee requirements. The best substitution was selected from the remaining substitutions. 
A static \emph{tuning parameter} ($\theta$) was introduced to demonstrate changes to the guarantees according to the degree they are inherited. 
The tuning parameter is insufficient to demonstrate changes to guarantees in a dynamic environment, but can represent effects of uncertainties. 
Larger tuning parameters signify a higher degree of inheritance and is analogous to when there are fewer uncertainties. 
The inherited guarantee establishes the requirement that substitutions must meet. 

\subsection{Experiments}
All approaches were evaluated against models described in \cite{8387659}. 
The modelled \emph{rover} requires positional information for the rover with respect to an obstacle, and the modelled \emph{drivetrain} requires positional information for a steered wheel axle.  
Balanced and random trees were used to assess knowledge bases of varying sizes. They included a balanced tree with a branching factor of 2 and a depth of 8 while each leaf node has equal properties. Balanced trees with a fixed branching factor of 2 at increasing depth with properties randomly assigned to leaf nodes were also included. This allowed observations to be made on the predictability of knowledge bases consisting of nodes with various properties.
The tuning parameter was adjusted to increase from 0.01 to 0.99. It is possible to not return a substitution if guarantee requirements were not satisfied. 
Experiments also tried to identify the lower and upper boundaries that may limit the pool of valid substitutions for some guarantee requirements. The size of the pool also influences the time required to find valid substitutions.

\begin{table*}[htbp]
%\tiny
% \scriptsize
\caption{Average time (ms) taken to search the knowledge bases and return a number of valid substitutions and satisfies guarantees with respect to tuning parameters 0.9 , 0.95, and 0.99. The balance tree has a branching factor of 2 and depth of 8.}
\begin{center}
\begin{tabular}{|M{0.06\linewidth}|M{0.035\linewidth}|M{0.055\linewidth}|M{0.035\linewidth}|M{0.055\linewidth}|M{0.035\linewidth}|M{0.055\linewidth}|M{0.06\linewidth}|M{0.035\linewidth}|M{0.055\linewidth}|M{0.035\linewidth}|M{0.055\linewidth}|M{0.035\linewidth}|M{0.055\linewidth}|}
\hline
\multirow{2}{*}{\parbox{\linewidth}{\centering{Model}}} & \multicolumn{2}{c|}{ORR} & \multicolumn{2}{c|}{SHPGSA} & \multicolumn{2}{c|}{DFS} & \multirow{2}{*}{\parbox{\linewidth}{\centering{Tune ($\theta$)}}} & \multicolumn{2}{c|}{ORRG} & \multicolumn{2}{c|}{SHPGSAG} & \multicolumn{2}{c|}{DFSG} \\
\cline{2-7}\cline{9-14} & subs & time & subs & time & subs & time & & subs & time & subs & time & subs & time \\
\hline
\multirow{3}{*}{\parbox{\linewidth}{\centering{rover}}} & \multirow{3}{*}{\parbox{\linewidth}{\centering{1}}} & \multirow{3}{*}{\parbox{\linewidth}{\centering{0.016}}} & \multirow{3}{*}{\parbox{\linewidth}{\centering{1}}} & \multirow{3}{*}{\parbox{\linewidth}{\centering{0.089}}} & \multirow{3}{*}{\parbox{\linewidth}{\centering{3}}} & \multirow{3}{*}{\parbox{\linewidth}{\centering{0.186}}} & 0.9 & 1 & 0.209 & 1 & 0.093 & 1 & 0.332 \\
\cline{8-14} & & & & & & & 0.95 & 1 & 0.133 & 1 & 0.095 & 1 & 0.348 \\
\cline{8-14} & & & & & & & 0.99 & 0 & 0.220 & 0 & 0.056 & 0 & 0.346 \\
\hline
\multirow{3}{*}{\parbox{\linewidth}{\centering{drivetrain}}} & \multirow{3}{*}{\parbox{\linewidth}{\centering{1}}} & \multirow{3}{*}{\parbox{\linewidth}{\centering{0.021}}} & \multirow{3}{*}{\parbox{\linewidth}{\centering{1}}} & \multirow{3}{*}{\parbox{\linewidth}{\centering{0.124}}} & \multirow{3}{*}{\parbox{\linewidth}{\centering{2}}} & \multirow{3}{*}{\parbox{\linewidth}{\centering{0.394}}} & 0.9 & 1 & 0.142 & 1 & 0.125 & 1 & 1.002 \\
\cline{8-14} & & & & & & & 0.95 & 0 & 0.297 & 0 & 0.123 & 0 & 0.518 \\
\cline{8-14} & & & & & & & 0.99 & 0 & 0.288 & 0 & 0.125 & 0 & 0.296 \\
\hline
\multirow{3}{*}{\parbox{\linewidth}{\centering{balanced}}} & \multirow{3}{*}{\parbox{\linewidth}{\centering{1}}} & \multirow{3}{*}{\parbox{\linewidth}{\centering{0.069}}} & \multirow{3}{*}{\parbox{\linewidth}{\centering{1}}} & \multirow{3}{*}{\parbox{\linewidth}{\centering{445.141}}} & \multirow{3}{*}{\parbox{\linewidth}{\centering{32768}}} & \multirow{3}{*}{\parbox{\linewidth}{\centering{8435.086}}} & 0.9 & 0 & 2.437 & 0 & 13.86 & 0 & 354.029 \\
\cline{8-14} & & & & & & & 0.95 & 0 & 2.433 & 0 & 3.651 & 0 & 186.664 \\
\cline{8-14} & & & & & & & 0.99 & 0 & 2.441 & 0 & 1.884 & 0 & 13.338 \\
\hline
\end{tabular}
\label{tab:a1} 
\end{center}
\end{table*}

\begin{figure}[htbp]
\centering
\includegraphics[height=4cm]{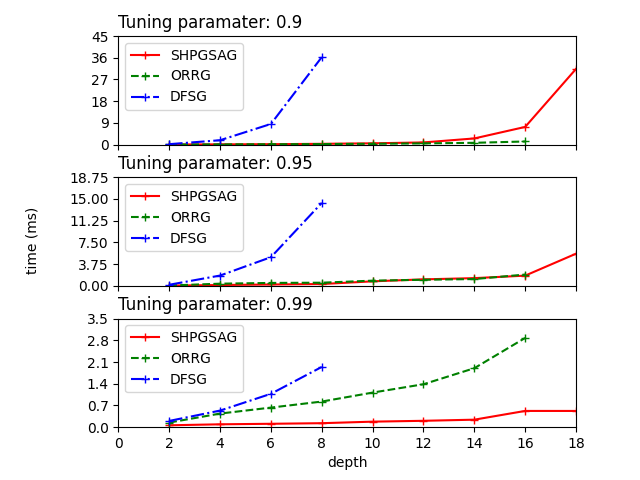}
\caption{Average time (s) taken to search for valid substitutions from random tree knowledge bases with branching factor of 2 and increasing depth.}
\label{fig:exG909599}
\end{figure}

\begin{figure}[htbp]
\centering
\includegraphics[height=4cm]{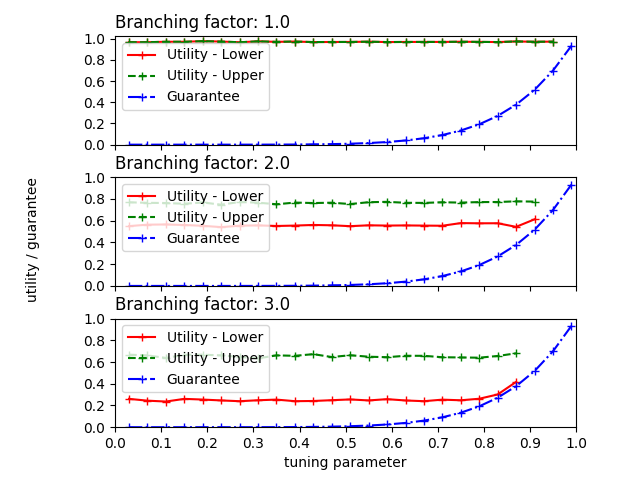}
\caption{Limiting substitutions against increasing tuning parameters. Upper and lower utility outlines the pool of valid substitutions.}
\label{fig:exUtilLim}
\end{figure}

\begin{figure}[htbp]
\centering
\includegraphics[height=1.7cm]{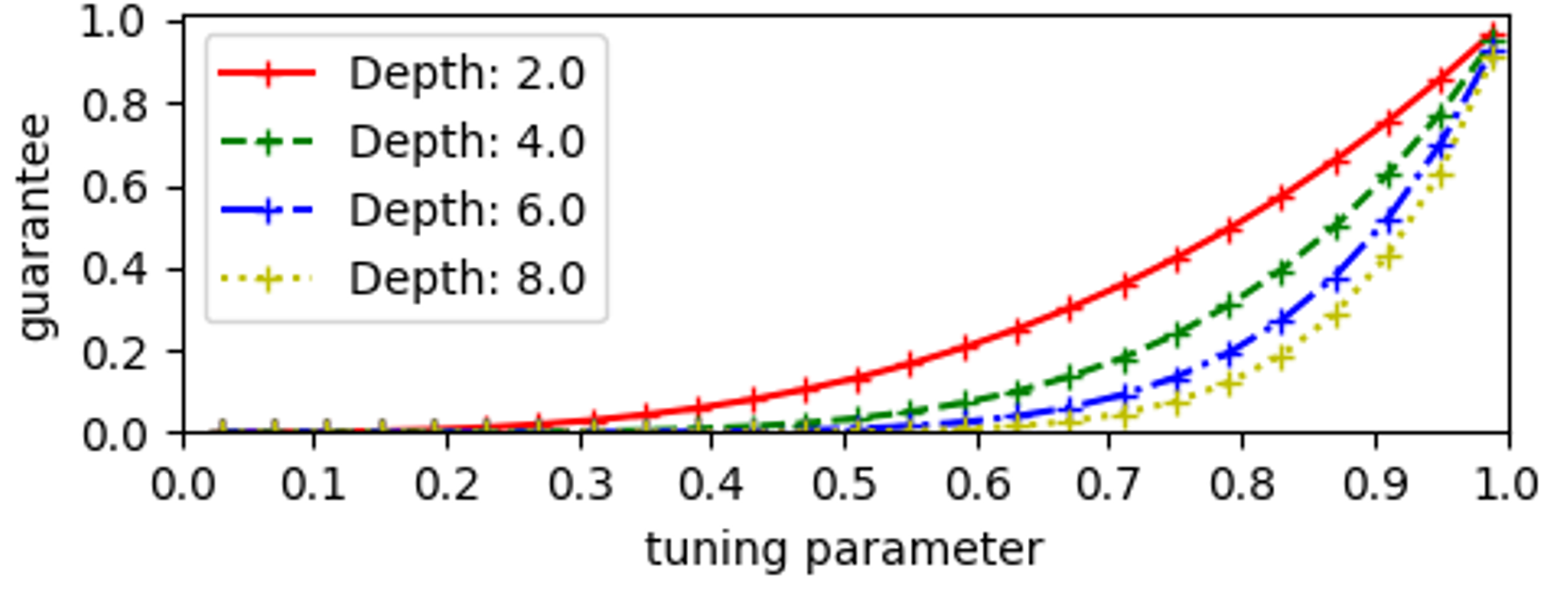}
\caption{Distribution of guarantees with respect to tuning parameters for balanced tree knowledge bases with branching factor of 3.}
\label{fig:exGua2Dep}
\end{figure}

\subsection{Analysis}
Table~\ref{tab:a1} shows attempts for each approach to return valid substitutions for guarantees against some tuning parameter. Results include the number of valid substitutions and average time taken for each approach. 
The number of valid substitutions decreased as the tuning parameter increases guarantee requirements until no substitutions can provide the required guarantees (e.g. 0 substitutions were returned for tuning parameter of 0.99). 
DFSG took the longest time to find valid substations as it exhaustively search the knowledge base to find substitutions that met guarantee requirements. 
The time taken by the ORRG approach is very much dependent on its ability to quickly find a substitution that meets guarantee requirements during early iterations. 
Similar to DFSG, ORRG exhaustively searched the knowledge base until it found a substitution that was valid and met the guarantee requirements. 
Finally, SHPGSAG returned the substitution with the highest utility that satisfied the guarantee requirements. 
On average, SHPGSAG took the least time to evaluate whether substitutions met guarantee requirements as requirements increased. 
 Figure~\ref{fig:exG909599} illustrates the behaviour of SH approaches addressing guarantee requirements with respect to tuning parameters. Here, knowledge bases were modelled with random trees with a branching factor of 2 and increasing depth with properties randomly assigned to leaf nodes. DFSG took the longest to find valid substitutions while ORRG and SHPGSAG performed similarly for tuning parameters at 0.9 and 0.95. However, SHPGSAG outperforms ORRG and DFSG to determine that there were no valid substitutions when the tuning parameter was 0.99. 
It becomes difficult to find valid substitutions as the requirement for a guarantee increases. Figure~\ref{fig:exUtilLim} shows changes in the utility of a set of substitutions as the guarantee requirements increase with the tuning parameter. It indicates a decrease in the number of valid substitutions that guarantee a level of dependability. Finally, longer node sequences in substitutions further influences the guarantees as successive inheritances sharply decrease guarantees with respect to the tuning parameter as shown in Figure~\ref{fig:exGua2Dep}. 

%%%%%%%%%%%%%%%%%%%%%%%%%%%%%%%%%%%%%%%%%%%%%%%%%%%%%%%%%%%%%%%%%%%%%%%%
% discussion
\section{Discussion} \label{sect:5}
% hierarchical guarantees
SH-CPS should guarantee a level of dependability after recovering from faults and failures. This work addresses this by inheriting or decomposing guarantees from predecessor or successor nodes while observing uncertainties. 
Inheriting guarantees provide information on the level of dependability of predecessor nodes. 
A minimum level of dependability substitutions should ensure in the event of faults and failures can be established. Alternatively, the guarantee values may be decomposed from the successor nodes to establish the requirements that substitutions must fulfil for node higher in the hierarchy. This is important for the interoperability of composite services and systems as they require a level of QoS while information is communicated across various domains \cite{Wang2020ServiceCI}. 
An approach to automatically generate hierarchical guarantees by decomposing guarantees of a root system with its underlying components was shared in \cite{8864556}. This is suitable for subsystem functions where the guarantees of some nodes have already been established. 
The lengthy sequences of nodes also affects the cost of operations to assign guarantees. 
To traverse the entire knowledge base to assign guarantees, the maximum cost of operations is $O(N+E)$. 
However, the minimum cost of operations to assign guarantees is $O(N_{d+1}+E_{d+1})$ when the guarantees of the predecessor nodes are readily available. 

SH-CPS should also consider the trade-offs between efficiency, flexibility, and resilience \cite{ivanov:emse-00881470}. 
Although it is ideal to maximise functionality while minimising costs, there is a complexity for systems leveraging on redundacy. Various factors are associated with resilience through redundancy such as energy use and cost of quality of control \cite{XU2020100411}. 
While costs for explicitly redundant systems are inherently higher than implicit redundanct systems, implicit redundant systems should recognise the changes in functionality after recovery.
Both approaches have distint effects to the efficiency of the system in the short and long term \cite{10.1145/3549073}. 

\section{Threats to validity} \label{sect:6}
SH-CPS are exposed to uncertainties that arise from the complexity of the system and the unpredictability of its environment \cite{8606923,s21165397}. This work represents uncertainty in the form of a tuning parameter that concerns the transmission of functional information between multiple nodes. Data transmission along long sequences of nodes may be exposed to more uncertainties that affect the behaviour of SH-CPS, affecting guarantees \cite{Ma2021}. 
Several factors contribute to the uncertainty of CPS, including the lack of information and the inherent variability \cite{10.1007/978-3-319-42061-5_16}. As such, assigning static values to tuning parameters does not fully reflect the intricacies of SH-CPS. 

%%%%%%%%%%%%%%%%%%%%%%%%%%%%%%%%%%%%%%%%%%%%%%%%%%%%%%%%%%%%%%%%%%%%%%%%
% conclude and future work
\section{Conclusion} \label{sect:7}
SH-CPS should guarantee a level of dependability when recovering from system perceived faults and failures. Implicitly redundant SH-CPS leverages implicitly redundant components to recover from a failure through structural adaptation requires additional mechanisms to transform data as required. These transformations do not necessarily guarantee dependability. Our work introduces implicit guarantees to ensure a level of dependability in implicitly redundant SH-CPS, where implicitly redundant components inherits or decomposes guarantees from predecessor or successor components. The experiments demonstrated the importance of implicit guarantees when searching and selecting adaptive solutions to address the guarantee requirements. 
Decomposing guarantees can be explored in more detail in situations where more than one predecessor is involved in future work. Guarantees in dynamic environments, where external factors influence properties, are also of interest.

%%%%%%%%%%%%%%%%%%%%%%%%%%%%%%%%%%%%%%%%%%%%%%%%%%%%%%%%%%%%%%%%%%%%%%%%

\printbibliography

@INPROCEEDINGS{8780540,  author={Ratasich, Denise and Platzer, Michael and Grosu, Radu and Bartocci, Ezio},  booktitle={2019 IEEE 13th International Conference on Self-Adaptive and Self-Organizing Systems (SASO)},   title={Adaptive Fault Detection Exploiting Redundancy with Uncertainties in Space and Time},   year={2019},  volume={},  number={},  pages={23-32},  doi={10.1109/SASO.2019.00013}}

@INPROCEEDINGS{8387659,  author={Ratasich, Denise and Preindl, Thomas and Selyunin, Konstantin and Grosu, Radu},  booktitle={2018 IEEE Industrial Cyber-Physical Systems (ICPS)},   title={Self-healing by property-guided structural adaptation},   year={2018},  volume={},  number={},  pages={199-205},  doi={10.1109/ICPHYS.2018.8387659}}

@INPROCEEDINGS{8864556,  author={Xu, Zhiheng and Ng, Daniel Jun Xian and Easwaran, Arvind},  booktitle={2019 IEEE 25th International Conference on Embedded and Real-Time Computing Systems and Applications (RTCSA)},   title={Automatic Generation of Hierarchical Contracts for Resilience in Cyber-Physical Systems},   year={2019},  volume={},  number={},  pages={1-11},  doi={10.1109/RTCSA.2019.8864556}}

@ARTICLE{8606923,  author={Ratasich, Denise and Khalid, Faiq and Geissler, Florian and Grosu, Radu and Shafique, Muhammad and Bartocci, Ezio},  journal={IEEE Access},   title={A Roadmap Toward the Resilient Internet of Things for Cyber-Physical Systems},   year={2019},  volume={7},  number={},  pages={13260-13283},  doi={10.1109/ACCESS.2019.2891969}}

@article{GHOSH20072164,
title = {Self-healing systems — survey and synthesis},
journal = {Decision Support Systems},
volume = {42},
number = {4},
pages = {2164-2185},
year = {2007},
note = {Decision Support Systems in Emerging Economies},
issn = {0167-9236},
doi = {https://doi.org/10.1016/j.dss.2006.06.011},
%%url = {https://www.sciencedirect.com/science/article/pii/S0167923606000807},
author = {Debanjan Ghosh and Raj Sharman and H. {Raghav Rao} and Shambhu Upadhyaya},
keywords = {Software engineering designing, Software architecture, Fault tolerance, Self-healing, Decision support systems, Distributed systems, Adaptive systems, Survivable systems},
abstract = {As modern software-based systems and applications gain in versatility and functionality, the ability to manage inconsistent resources and service disparate user requirements becomes increasingly imperative. Furthermore, as systems increase in complexity, rectification of system faults and recovery from malicious attacks become more difficult, labor-intensive, expensive, and error-prone. These factors have actuated research dealing with the concept of self-healing systems. Self-healing systems attempt to “heal” themselves in the sense of recovering from faults and regaining normative performance levels independently the concept derives from the manner in which a biological system heals a wound. Such systems employ models, whether external or internal, to monitor system behavior and use inputs obtaining therefore to adapt themselves to the run-time environment. Researchers have approached this concept from several different angles this paper surveys research in this field and proposes a strategy of synthesis and classification.}
}

@INPROCEEDINGS{7194653,  author={Arcaini, Paolo and Riccobene, Elvinia and Scandurra, Patrizia},  booktitle={2015 IEEE/ACM 10th International Symposium on Software Engineering for Adaptive and Self-Managing Systems},   title={Modeling and Analyzing MAPE-K Feedback Loops for Self-Adaptation},   year={2015},  volume={},  number={},  pages={13-23},  doi={10.1109/SEAMS.2015.10}}

@ARTICLE{1160055,  author={Kephart, J.O. and Chess, D.M.},  journal={Computer},   title={The vision of autonomic computing},   year={2003},  volume={36},  number={1},  pages={41-50},  doi={10.1109/MC.2003.1160055}}

@article{Ma2018ModelingFF,
  title={Modeling foundations for executable model-based testing of self-healing cyber-physical systems},
  author={Tao Ma and Shaukat Ali and Tao Yue},
  journal={Software \& Systems Modeling},
  year={2018},
  pages={1-31}
}

@article{Nelson1990FaulttolerantCF,
  title={Fault-tolerant computing: fundamental concepts},
  author={Victor P. Nelson},
  journal={Computer},
  year={1990},
  volume={23},
  pages={19-25}
}

@ARTICLE{1432851,  author={Seung-Chul Lee and Chan-Eom Park},  journal={IEEE Transactions on Energy Conversion},   title={Sensor value validation based on implicit sensor redundancy for reliable operation of power plants},   year={2005},  volume={20},  number={2},  pages={373-380},  doi={10.1109/TEC.2004.841521}}

@article{KHALIL2019104620,
title = {Self-healing hardware systems: A review},
journal = {Microelectronics Journal},
volume = {93},
pages = {104620},
year = {2019},
issn = {0026-2692},
doi = {https://doi.org/10.1016/j.mejo.2019.104620},
%url = {https://www.sciencedirect.com/science/article/pii/S0026269219302782},
author = {Kasem Khalil and Omar Eldash and Ashok Kumar and Magdy Bayoumi},
abstract = {Self-healing is increasingly becoming a promising approach to designing reliable digital systems, and it refers to the ability of a system to detect faults or failures and fix them through healing or repairing. Digital systems with architecture for self-healing are expected to compensate faults. However, there are a few research challenges that need to be overcome before self-healing becomes a mainstream approach. For example, current self-healing techniques face challenges such as scalability, reliability, area overhead, and mapping. This paper explains the self-healing concept and investigates the self-healing approaches related to digital design in the literature. It gives a general overview of the topic and explains levels of abstraction at which self-healing can be used: hardware level, application level, and system level. The paper presents multiple related works at each level of abstraction. In the faulted phase, different types of faults and fault detection methods are described. For the evaluation of a self-healing technique, this paper presents the parameters which can be used to evaluate a self-healing method. These parameters are redundancy rate, the maximum ratio of repair, self-healing time consumption, reliability, and area overhead. The paper also presents a comparison between previous works of self-healing in terms of evaluation techniques. Implementations using VHDL and ISE Xilinx Vertex-5 for self-healing on Embryonic Hardware (EmHW) and Network-on-Chip (NoC) are also presented. The simulation results show the contribution of self-healing to improve system reliability and mean time to failure.}
}

@article{Miclea2011AboutDI,
  title={About dependability in cyber-physical systems},
  author={Liviu C. Miclea and Teodora Sanislav},
  journal={2011 9th East-West Design \& Test Symposium (EWDTS)},
  year={2011},
  pages={17-21}
}

@INPROCEEDINGS{5523280,  author={Rajkumar, Ragunathan and Lee, Insup and Sha, Lui and Stankovic, John},  booktitle={Design Automation Conference},   title={Cyber-physical systems: The next computing revolution},   year={2010},  volume={},  number={},  pages={731-736},  doi={10.1145/1837274.1837461}}

@Article{s19051033,
AUTHOR = {Zhou, Peng and Zuo, Decheng and Hou, Kun Mean and Zhang, Zhan and Dong, Jian and Li, Jianjin and Zhou, Haiying},
TITLE = {A Comprehensive Technological Survey on the Dependable Self-Management CPS: From Self-Adaptive Architecture to Self-Management Strategies},
JOURNAL = {Sensors},
VOLUME = {19},
YEAR = {2019},
NUMBER = {5},
ARTICLE-NUMBER = {1033},
%url = {https://www.mdpi.com/1424-8220/19/5/1033},
ISSN = {1424-8220},
ABSTRACT = {Cyber Physical Systems (CPS) has been a popular research area in the last decade. The dependability of CPS is still a critical issue, and few surveys have been published in this domain. CPS is a dynamic complex system, which involves various multidisciplinary technologies. To avoid human errors and to simplify management, self-management CPS (SCPS) is a wise choice. To achieve dependable self-management, systematic solutions are necessary to verify the design and to guarantee the safety of self-adaptation decisions, as well as to maintain the health of SCPS. This survey first recalls the concepts of dependability, and proposes a generic environment-in-loop processing flow of self-management CPS, and then analyzes the error sources and challenges of self-management through the formal feedback flow. Focusing on reducing the complexity, we first survey the self-adaptive architecture approaches and applied dependability means, then we introduce a hybrid multi-role self-adaptive architecture, and discuss the supporting technologies for dependable self-management at the architecture level. Focus on dependable environment-centered adaptation, we investigate the verification and validation (V&amp;V) methods for making safe self-adaptation decision and the solutions for processing decision dependably. For system-centered adaptation, the comprehensive self-healing methods are summarized. Finally, we analyze the missing pieces of the technology puzzle and the future directions. In this survey, the technical trends for dependable CPS design and maintenance are discussed, an all-in-one solution is proposed to integrate these technologies and build a dependable organic SCPS. To the best of our knowledge, this is the first comprehensive survey on dependable SCPS building and evaluation.},
DOI = {10.3390/s19051033}
}

@InProceedings{10.1007/978-3-319-42061-5_16,
author="Zhang, Man
and Selic, Bran
and Ali, Shaukat
and Yue, Tao
and Okariz, Oscar
and Norgren, Roland",
editor="W{\k{a}}sowski, Andrzej
and L{\"o}nn, Henrik",
title="Understanding Uncertainty in Cyber-Physical Systems: A Conceptual Model",
booktitle="Modelling Foundations and Applications",
year="2016",
publisher="Springer International Publishing",
address="Cham",
pages="247--264",
isbn="978-3-319-42061-5"
}

@Article{Zhang2019,
author={Zhang, Man
and Ali, Shaukat
and Yue, Tao
and Norgren, Roland
and Okariz, Oscar},
title={Uncertainty-Wise Cyber-Physical System test modeling},
journal={Software {\&} Systems Modeling},
year={2019},
month=Apr,
day={01},
volume={18},
number={2},
pages={1379-1418},
abstract={It is important that a Cyber-Physical System (CPS) with uncertainty in its behavior caused by its unpredictable operating environment, to ensure its reliable operation. One method to ensure that the CPS will handle such uncertainty during its operation is by testing the CPS with model-based testing (MBT) techniques. However, existing MBT techniques do not explicitly capture uncertainty in test ready models, i.e., capturing the uncertain expected behavior of a CPS in the presence of environment uncertainty. To fill this gap, we present an Uncertainty-Wise test-modeling framework, named as UncerTum, to create test ready models to support MBT of CPSs facing uncertainty. UncerTum relies on the definition of a UML profile [the UML Uncertainty Profile (UUP)] and a set of UML Model Libraries extending the UML profile for Modeling and Analysis of Real-Time and Embedded Systems (MARTE). UncerTum also benefits from the UML Testing Profile V.2 to support standard-based MBT. UncerTum was evaluated with two industrial CPS case studies, one real-world case study, and one open-source CPS case study from the following four perspectives: (1) Completeness and Coverage of the profiles and Model Libraries in terms of concepts defined in their underlying uncertainty conceptual model for CPSs, i.e., U-Model and MARTE, (2) Effort required to model uncertainty with UncerTum, and (3) Correctness of the developed test ready models, which was assessed via model execution. Based on the evaluation, we can conclude that we were successful in modeling all the uncertainties identified in the four case studies, which gives us an indication that UncerTum is sufficiently complete. In terms of modeling effort, we concluded that on average UncerTum requires 18.5{\%} more time to apply stereotypes from UUP on test ready models.},
issn={1619-1374},
doi={10.1007/s10270-017-0609-6},
%url={https://doi.org/10.1007/s10270-017-0609-6}
}

@ARTICLE{9466159,  author={Nikolić, Jovan and Jubatyrov, Nursultan and Pournaras, Evangelos},  journal={IEEE Transactions on Network and Service Management},   title={Self-Healing Dilemmas in Distributed Systems: Fault Correction vs. Fault Tolerance},   year={2021},  volume={18},  number={3},  pages={2728-2741},  doi={10.1109/TNSM.2021.3092939}}

@InProceedings{10.1007/978-3-319-58469-0_4,
author="Denzel, Michael
and Ryan, Mark
and Ritter, Eike",
editor="De Capitani di Vimercati, Sabrina
and Martinelli, Fabio",
title="A Malware-Tolerant, Self-Healing Industrial Control System Framework",
booktitle="ICT Systems Security and Privacy Protection",
year="2017",
publisher="Springer International Publishing",
address="Cham",
pages="46--60",
isbn="978-3-319-58469-0"
}

@Inbook{Cheng2009,
author="Cheng, Betty H. C.
and de Lemos, Rog{\'e}rio
and Giese, Holger
and Inverardi, Paola
and Magee, Jeff
and Andersson, Jesper
and Becker, Basil
and Bencomo, Nelly
and Brun, Yuriy
and Cukic, Bojan
and Di Marzo Serugendo, Giovanna
and Dustdar, Schahram
and Finkelstein, Anthony
and Gacek, Cristina
and Geihs, Kurt
and Grassi, Vincenzo
and Karsai, Gabor
and Kienle, Holger M.
and Kramer, Jeff
and Litoiu, Marin
and Malek, Sam
and Mirandola, Raffaela
and M{\"u}ller, Hausi A.
and Park, Sooyong
and Shaw, Mary
and Tichy, Matthias
and Tivoli, Massimo
and Weyns, Danny
and Whittle, Jon",
editor="Cheng, Betty H. C.
and de Lemos, Rog{\'e}rio
and Giese, Holger
and Inverardi, Paola
and Magee, Jeff",
title="Software Engineering for Self-Adaptive Systems: A Research Roadmap",
bookTitle="Software Engineering for Self-Adaptive Systems",
year="2009",
publisher="Springer Berlin Heidelberg",
address="Berlin, Heidelberg",
pages="1--26",
abstract="The goal of this roadmap paper is to summarize the state-of-the-art and to identify critical challenges for the systematic software engineering of self-adaptive systems. The paper is partitioned into four parts, one for each of the identified essential views of self-adaptation: modelling dimensions, requirements, engineering, and assurances. For each view, we present the state-of-the-art and the challenges that our community must address. This roadmap paper is a result of the Dagstuhl Seminar 08031 on ``Software Engineering for Self-Adaptive Systems,'' which took place in January 2008.",
isbn="978-3-642-02161-9",
doi="10.1007/978-3-642-02161-9_1",
%url="https://doi.org/10.1007/978-3-642-02161-9_1"
}

@INPROCEEDINGS{6913205,  author={Höftberger, Oliver and Obermaisser, Roman},  booktitle={16th IEEE International Symposium on Object/component/service-oriented Real-time distributed Computing (ISORC 2013)},   title={Ontology-based runtime reconfiguration of distributed embedded real-time systems},   year={2013},  volume={},  number={},  pages={1-9},  doi={10.1109/ISORC.2013.6913205}}

@article{Rajput2021MultiagentAF,
  title={Multi-agent architecture for fault recovery in self-healing systems},
  author={Pushpendra Kumar Rajput and Geeta Sikka},
  journal={Journal of Ambient Intelligence and Humanized Computing},
  year={2021},
  volume={12},
  pages={2849-2866}
}

@article{Abdi2018GuaranteedPS,
  title={Guaranteed Physical Security with Restart-Based Design for Cyber-Physical Systems},
  author={Fardin Abdi and Chien-Ying Chen and Monowar Hasan and Songran Liu and Sibin Mohan and Marco Caccamo},
  journal={2018 ACM/IEEE 9th International Conference on Cyber-Physical Systems (ICCPS)},
  year={2018},
  pages={10-21}
}

@article{Wang2020ServiceCI,
  title={Service Composition in Cyber-Physical-Social Systems},
  author={Shangguang Wang and Ao Zhou and Mingzhe Yang and Lei Sun and Ching-Hsien Hsu and Fangchun Yang},
  journal={IEEE Transactions on Emerging Topics in Computing},
  year={2020},
  volume={8},
  pages={82-91}
}

@article{Ma2021,
author={Ma, Tao
and Ali, Shaukat
and Yue, Tao},
title={Testing self-healing cyber-physical systems under uncertainty with reinforcement learning: an empirical study},
journal={Empirical Software Engineering},
year={2021},
month={4},
day={01},
volume={26},
number={3},
pages={52},
abstract={Self-healing is becoming an essential feature of Cyber-Physical Systems (CPSs). CPSs with this feature are named Self-Healing CPSs (SH-CPSs). SH-CPSs detect and recover from errors caused by hardware or software faults at runtime and handle uncertainties arising from their interactions with environments. Therefore, it is critical to test if SH-CPSs can still behave as expected under uncertainties. By testing an SH-CPS in various conditions and learning from testing results, reinforcement learning algorithms can gradually optimize their testing policies and apply the policies to detect failures, i.e., cases that the SH-CPS fails to behave as expected. However, there is insufficient evidence to know which reinforcement learning algorithms perform the best in terms of testing SH-CPSs behaviors including their self-healing behaviors under uncertainties. To this end, we conducted an empirical study to evaluate the performance of 14 combinations of reinforcement learning algorithms, with two value function learning based methods for operation invocations and seven policy optimization based algorithms for introducing uncertainties. Experimental results reveal that the 14 combinations of the algorithms achieved similar coverage of system states and transitions, and the combination of Q-learning and Uncertainty Policy Optimization (UPO) detected the most failures among the 14 combinations. On average, the Q-Learning and UPO combination managed to discover two times more failures than the others. Meanwhile, the combination took 52{\%} less time to find a failure. Regarding scalability, the time and space costs of the value function learning based methods grow, as the number of states and transitions of the system under test increases. In contrast, increasing the system's complexity has little impact on policy optimization based algorithms.},
issn={1573-7616},
doi={10.1007/s10664-021-09941-z},
%url={https://doi.org/10.1007/s10664-021-09941-z}
}

@inproceedings{10.1145/3424771.3424804,
author = {Dias, Jo\~{a}o Pedro and Sousa, Tiago Boldt and Restivo, Andr\'{e} and Ferreira, Hugo Sereno},
title = {A Pattern-Language for Self-Healing Internet-of-Things Systems},
year = {2020},
isbn = {9781450377690},
publisher = {Association for Computing Machinery},
address = {New York, NY, USA},
%url = {https://doi.org/10.1145/3424771.3424804},
doi = {10.1145/3424771.3424804},
abstract = {Internet-of-Things systems are assemblies of highly-distributed and heterogeneous parts that, in orchestration, work to provide valuable services to end-users in many scenarios. These systems depend on the correct operation of sensors, actuators, and third-party services, and the failure of a single one can hinder the proper functioning of the whole system, making error detection and recovery of paramount importance, but often overlooked. By drawing inspiration from other research areas, such as cloud, embedded, and mission-critical systems, we present a set of patterns for self-healing IoT systems. We discuss how their implementation can improve system reliability by providing error detection, error recovery, and health mechanisms maintenance.},
booktitle = {Proceedings of the European Conference on Pattern Languages of Programs 2020},
articleno = {25},
numpages = {17},
keywords = {internet-of-things, fault-tolerance, self-healing, patterns},
location = {Virtual Event, Germany},
series = {EuroPLoP '20}
}

@Inbook{Musil2017,
author="Musil, Angelika
and Musil, Juergen
and Weyns, Danny
and Bures, Tomas
and Muccini, Henry
and Sharaf, Mohammad",
editor="Biffl, Stefan
and L{\"u}der, Arndt
and Gerhard, Detlef",
title="Patterns for Self-Adaptation in Cyber-Physical Systems",
bookTitle="Multi-Disciplinary Engineering for Cyber-Physical Production Systems: Data Models and Software Solutions for Handling Complex Engineering Projects",
year="2017",
publisher="Springer International Publishing",
address="Cham",
pages="331--368",
abstract="Engineering Cyber-Physical Systems (CPS) is challenging, as these systems have to handle uncertainty and change during operation. A typical approach to deal with uncertainty is enhancing the system with self-adaptation capabilities. However, realizing self-adaptation in CPS, and consequently also in Cyber-Physical Production Systems (CPPS) as a member of the CPS family, is particularly challenging due to the specific characteristics of these systems, including the seamless integration of computational and physical components, the inherent heterogeneity and large-scale of such systems, and their open-endedness.",
isbn="978-3-319-56345-9",
doi="10.1007/978-3-319-56345-9_13",
%url="https://doi.org/10.1007/978-3-319-56345-9_13"
}

@ARTICLE{5386835,  author={Ganek, A. G. and Corbi, T. A.},  journal={IBM Systems Journal},   title={The dawning of the autonomic computing era},   year={2003},  volume={42},  number={1},  pages={5-18},  doi={10.1147/sj.421.0005}}

@article{WONG2022106934,
title = {Self-adaptive systems: A systematic literature review across categories and domains},
journal = {Information and Software Technology},
volume = {148},
pages = {106934},
year = {2022},
issn = {0950-5849},
doi = {https://doi.org/10.1016/j.infsof.2022.106934},
%url = {https://www.sciencedirect.com/science/article/pii/S0950584922000854},
author = {Terence Wong and Markus Wagner and Christoph Treude},
keywords = {Self-adaptive systems, Literature review},
}

@article{weyns2017software,
  title={Software engineering of self-adaptive systems: an organised tour and future challenges},
  author={Weyns, Danny},
  journal={Chapter in Handbook of Software Engineering},
  pages={2},
  year={2017},
  publisher={Springer}
}

@ARTICLE{8008800,  author={Calinescu, Radu and Weyns, Danny and Gerasimou, Simos and Iftikhar, Muhammad Usman and Habli, Ibrahim and Kelly, Tim},  journal={IEEE Transactions on Software Engineering},   title={Engineering Trustworthy Self-Adaptive Software with Dynamic Assurance Cases},   year={2018},  volume={44},  number={11},  pages={1039-1069},  doi={10.1109/TSE.2017.2738640}}

@inproceedings{10.1145/3106237.3106247,
author = {Maggio, Martina and Papadopoulos, Alessandro Vittorio and Filieri, Antonio and Hoffmann, Henry},
title = {Automated Control of Multiple Software Goals Using Multiple Actuators},
year = {2017},
isbn = {9781450351058},
publisher = {Association for Computing Machinery},
address = {New York, NY, USA},
%url = {https://doi.org/10.1145/3106237.3106247},
doi = {10.1145/3106237.3106247},
pages = {373–384},
numpages = {12},
keywords = {Non-Functional Requirements, Adaptive Software, Control Theory, Dynamic Systems},
location = {Paderborn, Germany},
series = {ESEC/FSE 2017}
}

@inproceedings{10.1145/2950290.2950301,
author = {Shevtsov, Stepan and Weyns, Danny},
title = {Keep It SIMPLEX: Satisfying Multiple Goals with Guarantees in Control-Based Self-Adaptive Systems},
year = {2016},
isbn = {9781450342186},
publisher = {Association for Computing Machinery},
address = {New York, NY, USA},
%url = {https://doi.org/10.1145/2950290.2950301},
doi = {10.1145/2950290.2950301},
booktitle = {Proceedings of the 2016 24th ACM SIGSOFT International Symposium on Foundations of Software Engineering},
pages = {229–241},
numpages = {13},
keywords = {Self-adaptive system, control theory, multiple goals, simplex},
location = {Seattle, WA, USA},
series = {FSE 2016}
}

@Article{s21062140,
AUTHOR = {Yeong, De Jong and Velasco-Hernandez, Gustavo and Barry, John and Walsh, Joseph},
TITLE = {Sensor and Sensor Fusion Technology in Autonomous Vehicles: A Review},
JOURNAL = {Sensors},
VOLUME = {21},
YEAR = {2021},
NUMBER = {6},
ARTICLE-NUMBER = {2140},
%url = {https://www.mdpi.com/1424-8220/21/6/2140},
PubMedID = {33803889},
ISSN = {1424-8220},
ABSTRACT = {With the significant advancement of sensor and communication technology and the reliable application of obstacle detection techniques and algorithms, automated driving is becoming a pivotal technology that can revolutionize the future of transportation and mobility. Sensors are fundamental to the perception of vehicle surroundings in an automated driving system, and the use and performance of multiple integrated sensors can directly determine the safety and feasibility of automated driving vehicles. Sensor calibration is the foundation block of any autonomous system and its constituent sensors and must be performed correctly before sensor fusion and obstacle detection processes may be implemented. This paper evaluates the capabilities and the technical performance of sensors which are commonly employed in autonomous vehicles, primarily focusing on a large selection of vision cameras, LiDAR sensors, and radar sensors and the various conditions in which such sensors may operate in practice. We present an overview of the three primary categories of sensor calibration and review existing open-source calibration packages for multi-sensor calibration and their compatibility with numerous commercial sensors. We also summarize the three main approaches to sensor fusion and review current state-of-the-art multi-sensor fusion techniques and algorithms for object detection in autonomous driving applications. The current paper, therefore, provides an end-to-end review of the hardware and software methods required for sensor fusion object detection. We conclude by highlighting some of the challenges in the sensor fusion field and propose possible future research directions for automated driving systems.},
DOI = {10.3390/s21062140}
}

@Article{s21165397,
AUTHOR = {Vargas, Jorge and Alsweiss, Suleiman and Toker, Onur and Razdan, Rahul and Santos, Joshua},
TITLE = {An Overview of Autonomous Vehicles Sensors and Their Vulnerability to Weather Conditions},
JOURNAL = {Sensors},
VOLUME = {21},
YEAR = {2021},
NUMBER = {16},
ARTICLE-NUMBER = {5397},
%url = {https://www.mdpi.com/1424-8220/21/16/5397},
PubMedID = {34450839},
ISSN = {1424-8220},
ABSTRACT = {Autonomous vehicles (AVs) rely on various types of sensor technologies to perceive the environment and to make logical decisions based on the gathered information similar to humans. Under ideal operating conditions, the perception systems (sensors onboard AVs) provide enough information to enable autonomous transportation and mobility. In practice, there are still several challenges that can impede the AV sensors’ operability and, in turn, degrade their performance under more realistic conditions that actually occur in the physical world. This paper specifically addresses the effects of different weather conditions (precipitation, fog, lightning, etc.) on the perception systems of AVs. In this work, the most common types of AV sensors and communication modules are included, namely: RADAR, LiDAR, ultrasonic, camera, and global navigation satellite system (GNSS). A comprehensive overview of their physical fundamentals, electromagnetic spectrum, and principle of operation is used to quantify the effects of various weather conditions on the performance of the selected AV sensors. This quantification will lead to several advantages in the simulation world by creating more realistic scenarios and by properly fusing responses from AV sensors in any object identification model used in AVs in the physical world. Moreover, it will assist in selecting the appropriate fading or attenuation models to be used in any X-in-the-loop (XIL, e.g., hardware-in-the-loop, software-in-the-loop, etc.) type of experiments to test and validate the manner AVs perceive the surrounding environment under certain conditions.},
DOI = {10.3390/s21165397}
}

@online{turtlebot3,
  author = {ROBOTIS},
  title = {{ROBOTIS eManual}},
  YEAR = {nodate},
  url = {https://emanual.robotis.com},
  urldate = {2022-12-14}
}

@article{10.1145/3549073,
author = {Linkov, Igor and Ligo, Alexandre and Stoddard, Kelsey and Perez, Beatrice and Strelzoffx, Andrew and Bellini, Emanuele and Kott, Alexander},
title = {Cyber Efficiency and Cyber Resilience},
year = {2023},
issue_date = {April 2023},
publisher = {Association for Computing Machinery},
address = {New York, NY, USA},
volume = {66},
number = {4},
issn = {0001-0782},
% url = {https://doi.org/10.1145/3549073},
doi = {10.1145/3549073},
abstract = {Complementary objectives competing for the same resources.},
journal = {Commun. ACM},
month = {mar},
pages = {33–37},
numpages = {5}
}

@article{XU2020100411,
title = {Enhancing dependability and energy efficiency of cyber-physical systems bydynamic actuator derating},
journal = {Sustainable Computing: Informatics and Systems},
volume = {28},
pages = {100411},
year = {2020},
issn = {2210-5379},
doi = {https://doi.org/10.1016/j.suscom.2020.100411},
% url = {https://www.sciencedirect.com/science/article/pii/S2210537920301384},
author = {Shikang Xu and Israel Koren and C. Mani Krishna},
keywords = {Cyber-physical systems, Energy consumption, Adaptive fault-tolerance,Lifetime and dependability, Sustainable computing},
abstract = {Thermal issues in the cyber part of cyber-physical systems (CPSs) hasattracted considerable attention in recent years. Heat generation due to energyconsumption results in accelerated thermal damage to the processing units, reducing theirlifetime. Fault tolerance contributes to a large fraction of this thermal behavior in CPSsince it is implemented using redundant computations. This paper studies the use ofdynamic actuator derating (i.e., artificially limiting the maximum actuator output) forreducing the need to apply maximum redundancy. By targeting the use of fault-tolerance,we are able to obtain significant reductions in computer energy expenditure and thermalstress without lowering the reliability. This has beneficial effects on processorlifetime as well as the required energy storage.}
}

@article{ivanov:emse-00881470,
  TITLE = {{The Ripple effect in supply chains: trade-off 'efficiency-flexibility-resilience' in supply chain disruption management}},
  AUTHOR = {Ivanov, Dmitry and Sokolov, Boris and Dolgui, Alexandre},
  % URL = {https://hal-emse.ccsd.cnrs.fr/emse-00881470},
  JOURNAL = {{International Journal of Production Research}},
  PUBLISHER = {{Taylor \& Francis}},
  VOLUME = {52},
  NUMBER = {7},
  PAGES = {2154-2172},
  YEAR = {2014},
  DOI = {10.1080/00207543.2013.858836},
  KEYWORDS = {dynamics ; supply chain ; control ; resilience ; robustness ; disruption management ; event management ; quantitative analysis ; information technology ; Ripple effect},
  PDF = {https://hal-emse.ccsd.cnrs.fr/emse-00881470/file/Ivanov2013.pdf},
  HAL_ID = {emse-00881470},
  HAL_VERSION = {v1},
}

\end{document}